\documentclass[iop]{emulateapj}

\usepackage{hyperref}
\usepackage{graphicx}
\usepackage{natbib}
\usepackage{amsmath,amsthm,amssymb}
\usepackage{url}

\newcommand{\code}[1]{\texttt{#1}}

\def\d{\ifmmode {\,\operatorname{d}}\else
    {$\operatorname{d}$}\fi}

\def\spose#1{\hbox to 0pt{#1\hss}}

\def\WISE{\textit{WISE}}

\def\km{\operatorname{km}}
\def\second{\operatorname{s}}
\def\Mpc{\operatorname{Mpc}}
\def\kms{\ifmmode {\rm\,km\,s^{-1}}\else
    ${\rm\,km\,s^{-1}}$\fi}
\def\kmsMpc{\ifmmode {\rm\,km\,s^{-1}\,Mpc^{-1}}\else
    ${\rm\,km\,s^{-1}\,Mpc^{-1}}$\fi}

\def\msun{\ifmmode {\rm\,M_\odot}\else ${\rm\,M_\odot}$\fi}
\def\Msun{\ifmmode {\rm\,M_\odot}\else ${\rm\,M_\odot}$\fi}
\def\lsun{\ifmmode \operatorname{L_\odot}\else $\operatorname{L_\odot}$\fi}
\def\Lsun{\ifmmode \operatorname{L_\odot}\else $\operatorname{L_\odot}$\fi}
\def\rsun{\ifmmode {\rm\,R_\odot}\else ${\rm\,R_\odot}$\fi}
\def\Rsun{\ifmmode {\rm\,R_\odot}\else ${\rm\,R_\odot}$\fi}

\def\cm{{\rm\,cm}}
\def\cm3{\ifmmode {\rm\,cm^{-3}}\else ${\rm\,cm^{-3}}$\fi}

\def\ergps{\ifmmode {\rm\,erg\,s^{-1}}\else ${\rm\,erg\,s^{-1}}$\fi}
\def\ergpscm2{\ifmmode {\rm\,erg\,s^{-1}\,cm^{-2}}\else
    ${\rm\,erg\,s^{-1}\,cm^{-2}}$\fi}

\def\microJy{\operatorname{\mu Jy}}

\def\deg{\ifmmode {^{\circ}}\else {$^\circ$}\fi}
\def\degr{\ifmmode {^{\circ}}\else {$^\circ$}\fi}
\def\degs{\ifmmode {^{\circ}}\else {$^\circ$}\fi}

\def\micron{\operatorname{\mu m}}
\def\nm{\operatorname{nm}}

\def\h3Mpc{h^{-3}{\rm Mpc}^3}
\def\Ho{\ifmmode {\rm\,H_\circ}\else ${\rm\,H_\circ}$\fi}
\def\hnot{\ifmmode {\rm\,H_\circ}\else ${\rm\,H_\circ}$\fi}
\def\h0{\ifmmode {\rm\,H_\circ}\else ${\rm\,H_\circ}$\fi}
\def\hnotunit{\ifmmode {\rm\,km\,s^{-1}\,Mpc^{-1}}\else
    ${\rm\,km\,s^{-1}\,Mpc^{-1}}$\fi}
\def\qnot{\ifmmode {\rm\,q_\circ}\else ${\rm q_\circ}$\fi}
\def\q0{\ifmmode {\rm\,q_\circ}\else ${\rm q_\circ}$\fi}

\def\arcsec{\ifmmode {^{\prime\prime}}\else $^{\prime\prime}$\fi}
\def\asec{\ifmmode {^{\prime\prime}}\else $^{\prime\prime}$\fi}
\def\arcmin{\ifmmode {^{\prime}}\else $^{\prime}$\fi}
\def\amin{\ifmmode {^{\prime}}\else $^{\prime}$\fi}

\def\secper{\ifmmode \rlap.{^{s}}\else $\rlap{.}{^{s}} $\fi}
\def\minper{\ifmmode \rlap.{^{m}}\else $\rlap{.}{^m} $\fi}
\def\magper{\ifmmode \rlap.{^{m}}\else $\rlap{.}{^m} $\fi}
\def\arcsper{\ifmmode \rlap.{^{\prime\prime}}\else
    $\rlap.{^{\prime\prime}}$\fi}
\def\arcmper{\ifmmode \rlap.{^{\prime}}\else
    $\rlap.{^{\prime}}$\fi}
\def\spose#1{\hbox to 0pt{#1\hss}}
\def\simlt{\mathrel{\spose{\lower 3pt\hbox{$\mathchar"218$}}
     \raise 2.0pt\hbox{$\mathchar"13C$}}}
\def\simgt{\mathrel{\spose{\lower 3pt\hbox{$\mathchar"218$}}
     \raise 2.0pt\hbox{$\mathchar"13E$}}}

\def\apjl{{ApJ}}

\def\apjs{{ApJS}}

\def\apjref#1;#2;#3;#4 {\par\pp#1, {#2}, #3, #4 \par}

\shorttitle{$K$-corrections}
\shortauthors{Lake et al.}

\begin{document}

\title{$K$-corrections: an Examination of their Contribution to the Uncertainty of Luminosity Measurements}
\author{Sean~E.~Lake\altaffilmark{1}, E.~L.~Wright\altaffilmark{1}}

\altaffiltext{1}{Physics and Astronomy Department, University of California, Los Angeles, CA 90095-1547}

\email{lake@physics.ucla.edu}

\begin{abstract}
In this paper we provide formulae that can be used to determine the uncertainty contributed to a measurement by a $K$-correction and, thus, valuable information about which flux measurement will provide the most accurate $K$-corrected luminosity. All of this is done at the level of a Gaussian approximation of the statistics involved, that is, where the galaxies in question can be characterized by a mean spectral energy distribution (SED) and a covariance function (spectral 2-point function). This paper also includes approximations of the SED mean and covariance for galaxies, and the three common subclasses thereof, based on applying the templates from \cite{Assef:2010} to the objects in zCOSMOS bright 10k \citep{Lilly:2009} and photometry of the same field from \cite{Capak:2007}, \cite{Sanders:2007}, and the AllWISE source catalog.
\end{abstract}



\section{Introduction}
The $K$-correction was originally defined in the work of \cite{Humason:1956}. As initially defined, it was limited to filter transforms from an observer frame photometric filter to the same filter in the galaxy's rest frame. Later work generalized this concept to include transforms to other rest frame observations \citep[for example, ][]{Blanton:2003K}. There is a thorough summary of the state of the art of $K$-corrections in \cite{Hogg:2002}. $K$-corrections are primarily useful when a large number of objects need to be characterized and there is not sufficient data about all of them to fully specify the spectral energy distribution (SED) of each object, or when theoretical knowledge of the objects' SEDs are deficient. Put in other words, $K$-corrections are the correct approach to take when the uncertainty in the predictions of the theoretical model exceeds the uncertainty of performing a filter transformation on a small number of observations. What has been missing in the literature, thus far, is an objective specification of which observation frame filter to choose to perform this transformation when multiple close filters are available, or, even better, how to combine two or more filters to increase the signal to noise ratio (SNR) of the resulting measurement.

The answer to both of the questions above, how to combine and which filters to choose, must be informed by an approximation of the contribution of the $K$-correction process to the uncertainty of the corrected measurement. It is also important to consider how systematic differences between fluxes measured using different filters can add differing biases. In particular, the biases in photometric measurements taken at different wavelengths will usually vary because of wavelength dependent background and resolution effects. It is also important to consider whether the post $K$-correction measurements need to be statistically independent (for example, for the construction of color-magnitude diagrams). Assuming systematic consistency is desirable beyond maximizing the SNR of every individual measurement, the answer to the question of which filters to $K$-correct and combine using an inverse variance weighted average is whichever filters produce $K$-corrected quantities with sufficiently high combined SNR for the data set as a whole. Examples of this sort of consideration include: Sloan Digital Sky Survey (SDSS) $i$ band measurements have significantly higher SNR than $z$ band ones, so it may yield more precise results for the data set as a whole to $K$-correct from observer frame $i$ to rest frame $z$ than from $z$ to $z$, even though the $z$ to $z$ correction can be smaller for a large number of galaxies.

Including information about the uncertainty added by the $K$-correction offers an improvement on the present state in the literature where filters are often chosen for $K$-correction based only on nearness of filters, regardless of whether the $K$-correction would move the flux across a spectral break with a wide range of strengths in galaxies' SEDs (for example, the 4,000\,\AA\ break). 

The structure of this paper is as follows: Section~\ref{sec:thry} contains a short derivation of the propagation of errors level (Gaussian statistics) uncertainty in the $K$-correction, Section~\ref{sec:data} describes the data used to measure the SED covariance function on galaxies (overall, red, blue, and Active Galactic Nuclei [AGN]), and Section~\ref{sec:results} summarizes the results of the measurement.

The cosmology used in this paper is based on the WMAP 9 year $\Lambda$CDM cosmology \citep{Hinshaw:2013}\footnote{\url{http://lambda.gsfc.nasa.gov/product/map/dr5/params/lcdm_wmap9.cfm}}, with flatness imposed, yielding: $\Omega_M = 0.2793,\ \Omega_\Lambda = 1 - \Omega_M$, and $H_0 = 70 \km \second^{-1} \Mpc^{-1}$ (giving Hubble time $t_H = H_0^{-1} = 13.97\operatorname{Gyr}$, and  Hubble distance $D_H =  c t_H = 4.283 \operatorname{Gpc}$). All magnitudes quoted are in the AB system, unless otherwise stated.

\section{Theory} \label{sec:thry}
The general form of the $K$-correction, adapted from Equation~9 of \cite{Hogg:2002} by inverting a fraction and changing variables in an integral, used here is shown in Equation~\ref{eqn:thry:Kcorr}:
\begin{align}
	K_{\mathrm{ratio}} & = \frac{1}{1+z} \left(  \frac{\int \frac{\d \nu}{\nu} L_\nu(\nu)Q(\nu)}{\int \frac{\d \nu}{\nu} g^Q_\nu(\nu) Q(\nu)} \right) \nonumber \\
	&\hphantom{=}\times \left( \frac{\int \frac{\d \nu}{\nu} g^R_\nu(\nu) R(\nu)}{ \int \frac{\d \nu}{\nu} L_\nu(\nu) R\left(\frac{\nu}{1+z}\right) } \right) , \label{eqn:thry:Kcorr}
\end{align}
where $R(\nu)$ is the observer frame detector's relative response to a photon of frequency $\nu$ (the Relative Photon Response [RPR]), $g^R_\nu(\nu)$ is the spectral energy distribution (SED) of the standard/zero point source of the observer's instrument, $L_\nu(\nu)$ is the rest frame luminosity SED of the source, and $Q(\nu)$ is the RPR of the instrument being $K$-corrected to (often Q = R). The usual definition of the $K$-correction is in terms of magnitudes, and in that case $K = 2.5 \log_{10} (K_{\mathrm{ratio}})$. The content of Equation~\ref{eqn:thry:Kcorr} can be summarized, in the notation of functional calculus, as:
\begin{align}
	K_{\mathrm{ratio}} & = \frac{1}{1+z} \left( \frac{L_{Qe}[L_\nu]}{L_{Ro}[L_\nu]}\right), \label{eqn:thry:Ksumm}
\end{align}
where $e$ and $o$ are added to the subscripts to emphasize that they are calculated in emitted frame and observer frame, respectively.

Both $L_{Qe}[L_\nu]$ and $L_{Ro}[L_\nu]$ are what are known as `functionals' of $L_\nu$ - functions that map an entire function to the real numbers. In particular, they fit into the class of linear functionals that have the general form:
\begin{align}
	f[L_\nu] & = \int w(\nu)\, L_\nu(\nu) \d \nu.
\end{align}
As long as the function $w$ is non-negative, and therefore falls into the class of weighting functions, then the form and units that $w$ has dictates the interpretation of the functional $f$. If $w = 1$, then $f$ is the bolometric luminosity. If $w(\nu) \propto \delta(\nu - \nu')$, then $f$ is proportional to a spectral luminosity. Most commonly in astronomy the weighting function is a detector's response to a photon, $w(\nu) \propto R(\nu) / \nu$. The weight function can also be proportional to $r^{-2}$, the inverse square of the distance, in which case all of the aforementioned quantities are fluxes instead of luminosities. 

The important part of the previous paragraph, establishing notation aside, is that the linearity of $L_{Qe}$ and $L_{Re}$ combines with the form of Equation~\ref{eqn:thry:Ksumm} to make $K_{\mathrm{ratio}}$ completely independent of the normalization of the SED. For concreteness, we define the normalization luminosity and the normalized SED, respectively, in terms of $w_N(\nu)$ to be:
\begin{align}
	L_N & \equiv \int w_N(\nu)\, L_\nu(\nu) \d \nu,\ \mathrm{and}\nonumber \\
	\ell_\nu(\nu) & \equiv \frac{L_\nu(\nu)}{L_N}.
\end{align}

Because the $K$-correction in Equation~\ref{eqn:thry:Ksumm} is also a functional of the SED, it is necessary to adapt standard multi-dimensional propagation of errors to functional calculus to calculate the uncertainty in $K_{\mathrm{ratio}}$. In multiple dimensions the propagation of errors formula that relates the covariance of some quantities, $\vec{x}$, to a vector valued function of those quantities, $\vec{f}(\vec{x})$, is:
\begin{align}
	\operatorname{cov}(f_i, f_j) & =  \sum_{m, n} \frac{\partial f_i(\vec{x})}{\partial x_m} \frac{\partial f_j(\vec{x})}{\partial x_n} \operatorname{cov}(x_m, x_n). \label{eqn:thry:propcount}
\end{align}
Equation~\ref{eqn:thry:propcount} generalizes immediately to functional calculus in an obvious way:
\begin{align}
	\operatorname{cov}(f_i, f_j) & = \int \frac{\delta f_i[\ell_\nu]}{\delta \ell_\nu(\nu)} \frac{\delta f_j[\ell_\nu]}{\delta \ell_\nu(\nu')} \Sigma(\nu, \nu') \d \nu \d \nu', \label{eqn:thry:properrs}
\end{align}
where $\Sigma(\nu, \nu')$ is the two point function of normalized SEDs in the class of galaxies being $K$-corrected; symbolically,
\begin{align}
	\mu_\nu(\nu) & \equiv  \left\langle \ell_\nu(\nu) \right\rangle, \ \mathrm{and} \nonumber \\
	\Sigma(\nu, \nu') & = \left\langle \left(\ell_\nu(\nu) - \mu_\nu(\nu) \right) \left(\ell_\nu(\nu') - \mu_\nu(\nu) \right) \right\rangle, \label{eqn:thry:sigdef}
\end{align}
where $\mu_\nu(\nu)$ is the mean SED.

The formula in Eqution~\ref{eqn:thry:properrs} is more general than is actually required because all fluxes and luminosities are linear functions of the SED, not general ones. So, if a set of luminosities is defined by positive semi-definite weight functions, $L_i = \int w_i(\nu)\, L_\nu(\nu) \d \nu$, then:
\begin{align}
	\operatorname{cov}(L_i, L_j) & = L_N^2 \int w_i(\nu) w_j(\nu') \Sigma(\nu, \nu') \d \nu \d \nu'. \label{eqn:thry:covlumas}
\end{align}

All of the tools are in place to produce the covariance of multiple $K$-corrections using the propagation of errors formalism. First, the variance of a single $K$-correction is:
\begin{align}
	\operatorname{var}(K_{\mathrm{ratio}} ) & = K_{\mathrm{ratio}}^2  \left(\frac{\operatorname{var}(L_{Q})}{L_{Q}^2} + \frac{\operatorname{var}(L_{R})}{L_{R}^2} \right. \nonumber \\
	&\hphantom{=K_{\mathrm{ratio}}^2 (-} \left.- 2 \frac{\operatorname{cov}(L_{Q},\, L_{R})}{L_{Q}\, L_{R}} \right), \label{eqn:thry:singlvar}
\end{align}
where the variances and covariance are calculated by applying Equation~\ref{eqn:thry:covlumas}. If multiple quantities are being $K$-corrected, then covariance matrix among the $K$-corrections takes the form:
\begin{widetext}
\begin{align}
	\operatorname{cov}(K_i,\, K_j) & = K_i K_j \left( \frac{\operatorname{cov}(L_{Qi},\, L_{Qj})}{L_{Qi}\, L_{Qj}} - \frac{\operatorname{cov}(L_{Qi},\, L_{Rj})}{L_{Qi}\, L_{Rj}}  - \frac{\operatorname{cov}(L_{Ri},\, L_{Qj})}{L_{Ri}\, L_{Qj}} + \frac{\operatorname{cov}(L_{Ri},\, L_{Rj})}{L_{Ri}\, L_{Rj}} \right). \label{eqn:thry:fullcovar}
\end{align}
\end{widetext}
The spectral versions of Equations~\ref{eqn:thry:singlvar} and \ref{eqn:thry:fullcovar} are:
\begin{widetext}
\begin{align}
	\operatorname{var}(K_{\mathrm{ratio}} ) & = K_{\mathrm{ratio}}^2  \left(\frac{\Sigma(\nu_Q,\, \nu_Q)}{\ell_\nu(\nu_Q)^2} + \frac{\Sigma(\nu_R,\, \nu_R)}{\ell_\nu(\nu_R)^2} - 2 \frac{\Sigma(\nu_Q,\, \nu_R)}{\ell_\nu(\nu_Q)\, \ell_\nu(\nu_R)} \right),\ \mathrm{and} \\
	\operatorname{cov}(K_i,\, K_j) & = K_i K_j \left( \frac{\Sigma(\nu_{Qi},\, \nu_{Qj})}{\ell_\nu(\nu_{Qi})\, \ell_\nu(\nu_{Qj})} - \frac{\Sigma(\nu_{Qi},\, \nu_{Rj})}{\ell_\nu(\nu_{Qi})\, \ell_\nu(\nu_{Rj})}  - \frac{\Sigma(\nu_{Ri},\, \nu_{Qj})}{\ell_\nu(\nu_{Ri})\, \ell_\nu(\nu_{Qj})} + \frac{\Sigma(\nu_{Ri},\, \nu_{Rj})}{\ell_\nu(\nu_{Ri})\, \ell_\nu(\nu_{Rj})} \right), \label{eqn:thry:spectralcovar}
\end{align}
\end{widetext}
respectively. It's worth reinforcing that the luminosity used for normalization, $L_N$, must be the same for calculating $\ell_\nu(\nu)$ and $\Sigma(\nu,\,\nu')$, as is required for the covariance of $K$-corrections to be as independent of normalization as the $K$-correction itself is.

If either $Q$ or the observer frame $R$ are proportional to the function that defines the SED normalization luminosity, then the form of Equation~\ref{eqn:thry:singlvar} simplifies greatly:
\begin{align}
	\operatorname{var}(K_{\mathrm{ratio}} ) & = K_{\mathrm{ratio}}^2 \frac{\operatorname{var}(L)}{L_N^2},
\end{align}
with a further simplification when the luminosity $L$ is a spectral luminosity at the frequency $\nu$:
\begin{align}
	\operatorname{var}(K_{\mathrm{ratio}} ) & = K_{\mathrm{ratio}}^2 \Sigma(\nu,\nu).\label{eqn:thry:spectral2p4}
\end{align}
The units in Equation~\ref{eqn:thry:spectral2p4} look a little odd because it is being evaluated in the special case where  $K_{\mathrm{ratio}} \propto L_\nu / L_N = \ell_\nu(\nu)$, or its multiplicative inverse, and therefore $\ell_\nu(\nu)$ is unitless by construction, making $\Sigma(\nu,\,\nu')$ unitless also.

The reason for exploring the simplified versions of the variance of $K_{\mathrm{ratio}}$ is that it highlights the centrality of $\Sigma(\nu,\nu')$ to the considerations here. Because of this its properties and the process of measuring it merit closer examination. $\Sigma$ has the property, clear by inspection of its definition, that it is symmetric under interchange of frequencies $\Sigma(\nu, \nu') = \Sigma(\nu', \nu)$. Less obvious is that $\Sigma$ has nodal lines that originate from the fact that all of the SEDs have to satisfy the normalization condition defined for $\ell_\nu(\nu)$. If the normalization luminosity is defined by the function $w_N(\nu)$, then the conditions imposed on the SEDs and $\Sigma$, respectively, are:
\begin{align}
	1 & = \int \ell_\nu(\nu)\, w_N(\nu) \d \nu,\ \mathrm{and} \nonumber \\
	0 & = \int \Sigma(\nu, \nu')\, w_N(\nu') \d \nu'.
\end{align}
If the normalization luminosity is even approximately spectral compared to the standard deviation of galaxy SEDs around frequency $\nu_N$, this condition will produce sharp sign flips on the $\nu=\nu_N$ and $\nu'=\nu_N$ axes in graphs of the correlation coefficient, $\rho(\nu,\nu') = \Sigma(\nu,\nu') / \sqrt{\Sigma(\nu,\nu)\, \Sigma(\nu',\nu')}$.

As with any covariance, $\Sigma(\nu,\nu')$ can be measured by replacing the expectation brackets in the definition, Equation~\ref{eqn:thry:sigdef}, with bias corrected sample averages:
\begin{widetext}
\begin{align}
	\Sigma(\nu,\nu') & \approx \frac{1}{N-1} \left(\sum_{i=1}^N \ell_{\nu,i}(\nu)\, \ell_{\nu,i}(\nu') - \frac{1}{N} \left[\sum_{i=1}^N \ell_\nu(\nu)\right] \cdot \left[\sum_{i=1}^N \ell_\nu(\nu')\right] \right). \label{eqn:thry:sampav}
\end{align}
\end{widetext}
It is usually not practical, however, to measure $\Sigma$ using full spectra. In this common case, it is possible to approximate Equation~\ref{eqn:thry:sampav} by writing each SED as a linear combination of $n_T$ template spectra, $\ell_{\nu,i}(\nu) = \sum_{j=1}^{n_T} f_{ij} \ell_{\nu,j}(\nu)$, as were produced in, for example, \cite{Assef:2010} and \cite{Rieke:2009}. Note that the templates have to be scaled to match the normalization condition, and when this is done the coefficients will satisfy $\sum_{j=1}^{n_T} f_{ij} = 1$. In terms of the template approximation, Equation~\ref{eqn:thry:sampav} becomes:
\begin{align}
	\mu_j & \equiv \frac{1}{N} \sum_{i=1}^N f_{ij}, \nonumber \\
	\sigma^2_{jk} & \equiv \frac{1}{N-1} \sum_{i=1}^N f_{ij} f_{ik} - \frac{N}{N-1} \mu_j \mu_k,\ \mathrm{and} \nonumber\\
	\Sigma(\nu,\nu') & \approx \sum_{j,k=1}^{n_T} \sigma^2_{jk}\, \ell_{\nu,k}(\nu)\, \ell_{\nu,j}(\nu);\label{eqn:thry:sigtmp}
\end{align}
that is, the templates reduce the infinite dimensional covariance function to an $n_T \times n_T$ covariance matrix of the template fractions.

\section{Data and Observations} \label{sec:data}
\begin{figure}[htb]
	\begin{center}
	\includegraphics[width=0.48\textwidth]{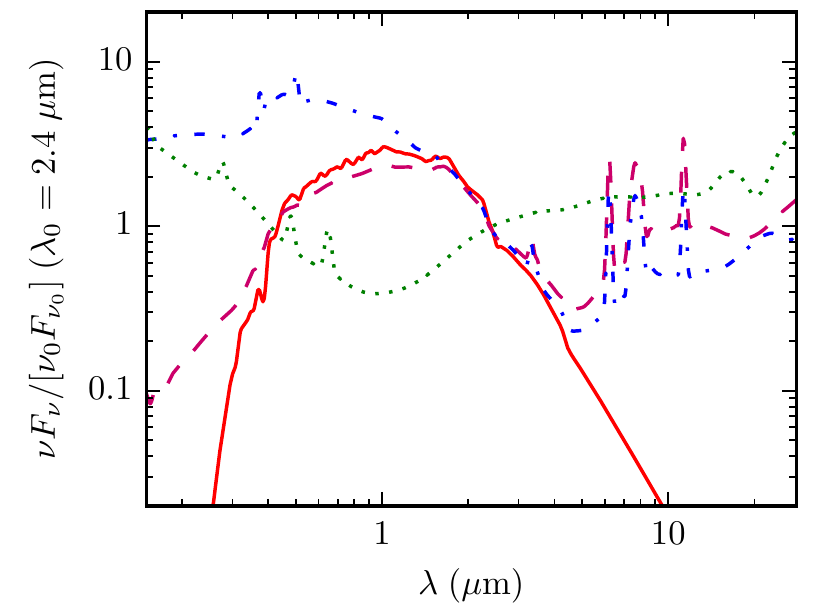}
	\end{center}
	
	\caption[\cite{Assef:2010} Template Spectra]{The template spectra used from \cite{Assef:2010}. 
		The red solid line is the template called ``Elliptical." The purple dashed line is ``Sbc." 
		The blue dash-dotted line is ``Irregular." And the green dotted line is 
		``AGN," unobscured. }
	\label{fig:dat:templates}
\end{figure}

The measurement of the galaxy SED covariance in this paper is based on the template spectra approximation outlined at the end of Section~\ref{sec:thry}. The template set used is the one defined in \cite{Assef:2010}. The set consists of four templates that correspond, roughly, to galaxies that are: red (named Elliptical), moderately star forming blue (Sbc), starburst blue (Irregular), and active galactic nuclei (AGN). The AGN template, additionally, has a dust obscuration model parametrized by $\operatorname{E}(B-V)$, the extinction excess. Graphs of the templates, normalized to the \WISE\ W1 filter at a redshift of $z=0.38$ (effective wavelength $\lambda \approx 2.4\micron$), can be found in Figure~\ref{fig:dat:templates}.  

The presence of AGN dust obscuration as a non-linear parameter throws off the mathematics behind Equations~\ref{eqn:thry:sigtmp}. There are multiple ways of getting around this problem, including replacing the AGN SED with multiple AGN SEDs that have different fixed extinction values. In order to make the results as simple as possible to produce, we only use one extinction value: the median $\operatorname{E}(B-V)$ for galaxies which had a sufficient AGN contribution to make the measured $\operatorname{E}(B-V)$ meaningful (see below). This means that the covariance measurement presented here will be an underestimate of the true spread among SEDs, particularly on the blue side of the spectrum. Estimating how much of an underestimate it is by looking at the distribution of $\operatorname{E}(B-V)$ values will prove inexact, because the effect of the parameter on individual SEDs is non-linear and the distribution of values observed is highly asymmetric. With those caveats in mind, we examined the distribution excess extinctions qualitatively and found it to have a width around 1 magnitude.

The scatter in observed AGN extinctions will be greater than what is imposed by dust near the black hole alone because galaxy inclination will change the amount of interstellar medium that the AGN's light has to travel through. Inclination will not just affect the measured AGN extinction, though, because the amount of dust the stellar light must travel through is also inclination dependent. The model we use in this work does not make any allowance for extinction within the target galaxy of anything but the AGN, though, so inclination effects will similarly modify the template fractions assigned to the galaxies in the fitting process. All of this has the effect of increasing the scatter of observed SEDs compared to the scatter that would be exhibited by a measure of the underlying physical properties of the galaxies. Despite these limitations, the work presented here should be sufficiently accurate for measuring a luminosity function in the near to mid-IR because of reduced dust absorption in those wavelengths.

Measuring the $f_{ij}$ of Equations~\ref{eqn:thry:sigtmp} using the \cite{Assef:2010} templates requires fitting the templates to observed photometry of a collection of galaxies, preferably with spectroscopic redshifts and a rich collection of filters. The zCOSMOS Bright 10k sample, described in \cite{Lilly:2009} and \cite{Knobel:2012}, is in the COSMOS field and, therefore, has a very rich set of publicly available photometry. The photometry we used is summarized in Table~\ref{tbl:dat:filters}. The targeting for the survey is based on photometry from Hubble Advanced Camera for Surveys (ACS) Wide Field Camera (WFC) imaging with the F814W filter, which is approximately $I$-band. The version of the data used for this analysis is Data Release 2.

The photometric surveys were cross-matched to the zCOSMOS data set based on a spatial cross-match that uniquely assigns a detection to its closest companion in zCOSMOS up to a maximum search radius that depended on the resolution of the external survey. For most surveys, the search radius was $1\arcsec$, but for AllWISE it was $3\arcsec$ (half the full width at half maximum of point sources for the \WISE\ W1 beam).

\begin{deluxetable*}{ccc}
	\tablewidth{0.75\textwidth}
	\tablecaption{zCOSMOS Photometry Used}
	\tablehead{\colhead{Survey} & \colhead{Bands} & \colhead{Citation} }
	\startdata
	COSMOS & FUV, NUV, $u^*$, $B_j$, $g^+$, $V_j$, \ldots &  \\
	& \ldots $r^+$, F814W, $i^+$, $i^*$, $z^+$, $J$, $K_s$ &  \cite{Capak:2007} \\
	SDSS-DR10 & $u$, $g$, $r$, $i$, $z$ & \cite{SDSSdr10} \\
	S-COSMOS-DR3 & c1, c2, c3, c4 & \cite{Sanders:2007} \\
	AllWISE & W1, W2, W3, W4 & \cite{Wright:2010} 
	\enddata
	\tablecomments{Photometric surveys used for fitting zCOSMOS sources. }
	\label{tbl:dat:filters}
\end{deluxetable*}

Selecting high quality redshifts from zCOSMOS is somewhat involved because of the detailed `confidence class' (\code{cc}) system used. The recommendation in \cite{Lilly:2009} is to accept all sources with \code{cc} equal to: any 3.X, 4.X, 1.5, 2.4, 2.5, 9.3, and 9.5. Based on the description of those classes, the analysis here accepted sources that fit in the recommended classes, but also those with a leading 1 (10 was added to show broad line AGN), 18.3, 18.5 (both broad line AGN consistent with the photometric redshift), and rejected all secondary targets (2 in the tens or hundreds digit). This can be done by accepting sources for which the text string version of \code{cc} matches the regular expression ``\verb=([34]\..*)|([1289]\.5)|(2\.4)|([89]\.3)=" and doesn't match ``\verb=^2\d+\.=". Finally, the targets fell into three selection classes, column named \code{i}, and `unintended' sources are rejected by requiring $\code{i}>0$.

In addition to good redshifts, the sources needed to have a minimum amount quality of photometry available to make the template fitting reliable. To that end, we limited the analysis to sources that meet all the following conditions: redshifts satisfy $0.05 < z \le 1.0$, the sources have at least five high quality photometric measurements (the number of free parameters in the SED fit when unconstrained, description follows), and were measured in S-COSMOS to have a have $F_{\mathrm{c1}} \ge 5 \microJy$ (corresponds to an empirical SNR limit of about 30, about $22.15\operatorname{mag}$) using a $3\arcsec$ aperture in \textit{Spitzer}'s IRAC channel 1 ($\lambda_{\mathrm{eff}} \approx 3.6\micron$). The external photometric measurements were deemed to be of sufficient quality if the photometry was not flagged as contaminated in the survey, or otherwise marked as obviously invalid by being less than or equal to zero.

The templates were constructed to be fit to fluxes using $\chi^2$ applied to a linear combination of the template fluxes with non-negative coefficients, and a search in the 1-dimensional parameter space for the best AGN extinction excess, $\operatorname{E}(B-V) = (2.5 / \ln(10)) \cdot (\tau_B - \tau_V)$. That is, the model has the form:

\begin{align}
	F_\nu(\nu, z) & = a_\mathrm{E} F_\mathrm{E}(\nu, z) + a_\mathrm{S} F_\mathrm{S}(\nu, z) \nonumber \\
	&\hphantom{=} + a_\mathrm{I} F_\mathrm{I}(\nu, z) + a_\mathrm{A} F_\mathrm{A}(\nu, z, \tau_B - \tau_V), \\
	\chi^2 & = \sum_{i \in \{\mathrm{filters}\}} \left(\frac{F_{\mathrm{obs}\,i} - F_{\mathrm{mod}\,i}}{\sigma_i}\right), \label{eqn:sedchisqr}
\end{align}
with all $a_i \ge 0$, and $12 > \tau_B - \tau_V \ge 0 $. The $a_i$ were fit using the SciPy \code{optimize} package's routine \code{nnls} (quadratic programming for non-negative least squares), and $\tau_B - \tau_V$ were fit with the routine \code{brent} with fallback to \code{fmin} (Nelder-Meade simplex).

There is one modification to that procedure for the fits done for this paper. The templates do not include the ability to tune dust obscuration of the galaxy's stars, so a dusty starburst that has a detection in \WISE's $12\micron$ filter, W3, will often be best fit with a galaxy that is dominated by its Elliptical component (to satisfy optical redness) and a super-obscured AGN ($\tau_B - \tau_V > 12$) masquerading as the emission from the stellar dust component. The problem this creates is that it makes the SED fit the data more poorly in the most important range for the subsequent uses to which we intend to put this data, where $K$-corrections from observer frame W1 to $2.4\micron$ rest wavelength are performed. We used two techniques to work around this problem. First, we limited the excess in optical depth as $\tau_B - \tau_V \le 12$ (equivalently, $\operatorname{E}(B-V) \le 13.03$). Second, when the SED was badly modeled ($\chi^2 > \mathrm{max}(N_{\mathrm{df}}, 1) \times 100$) and unlikely to be an AGN ($W1 - W2 > 0.5\operatorname{Vega\ mag}$ with uncertainty, $\sigma_{W1-W2} < 0.2\operatorname{Vega\ mag}$), we used the best model with $\operatorname{E}(B-V) = 0$. The reduced $\chi^2$ criterion was determined by subjective empirical examination, and the color based selection was found in \cite{Assef:2013} to select low redshift AGN with 90\% completeness. Overall, $2,604$ galaxies were fit using the `alternate' fitting mode where the AGN template was set at $\operatorname{E}(B-V) = 0$ and $4,621$ were fit in the `main' fitting mode where the AGN extinction was allowed to vary. For the subsets the breakdown is: none of the $268$ AGN, $1,139$ of the $1,903$ Red, and $1,465$ of the $5,054$ Blue galaxies were fit in the alternate mode. 

Limiting the excess optical depth, $\tau_B - \tau_V$, to be non-negative introduces a bias to the parameter estimation of the individual galaxies. It is even physically possible for a source to appear bluer than expected if the line of sight is unobscured and dust clouds are reflecting excess blue light into it (that is, the line of sight contains significant contribution from reflection nebulae in the target). Even so, applying a negative optical depth excess to dust obscuration models is not likely to produce an accurate spectrum for reflection, and the magnitude of the negative excess doesn't have to be large to cause the estimate of the maximum redshift at which the galaxy could be observed to diverge.

There is a final detail involved in dealing with AGN obscuration measurements. The impact of changes in $\tau_B - \tau_V$ on the shape of the overall SED depends on what fraction of the luminosity the AGN contributes. If a minuscule fraction of the luminosity is contributed by the AGN, then the shape of the SED is insensitive to how obscured the AGN template is, rendering the value that the fitting process assigns to $\tau_B - \tau_V$ meaningless. It is, therefore, necessary when computing statistics involving $\tau_B - \tau_V$ to limit the sample to those galaxies for which the AGN's contribution to the shape of the SED is non-negligible. The cutoff used in this work, set arbitrarily, is that the fraction of $2.4\micron$ luminosity contributed must be greater than $0.1\%$. The cutoff is set low for two reasons: first, the shapes of the template spectra mean that the ability to measure extinction in the AGN template depends on both what other templates are present and which wavelengths were observed; and second, we prefer to make less aggressive cuts to the data when making them without making a rigorous exploration of their impact on the data.

The resulting data set contains $7,225$ galaxies. The template fit parameters of the data set are included in this work (at figshare.com\footnote{\url{https://figshare.com/articles/zCOSMOS_Template_Fractions_tbl_gz/3804210}} with doi:10.6084/m9.figshare.3804210) in gzipped\footnote{\url{https://www.gnu.org/software/gzip/}} IPAC Table format\footnote{\url{http://irsa.ipac.caltech.edu/applications/DDGEN/Doc/ipac\_tbl.html}}, an excerpt from which is in Table~\ref{tbl:dat:fitpars}. The normalization condition chosen for the templates is the luminosity \WISE's W1 filter would observe directly in a galaxy at redshift $z=0.38$ ($\lambda_{\mathrm{eff}} \approx 2.4\micron$), after the effect of AGN obscuration has been applied to the AGN template. The latter choice ensures that the template fractions sum to 1, and that each $f$ represents the fraction of $2.4\micron$ luminosity contributed by the corresponding component of the galaxy.

\begin{figure}[htb]
	\begin{center}
	\includegraphics[width=0.48\textwidth]{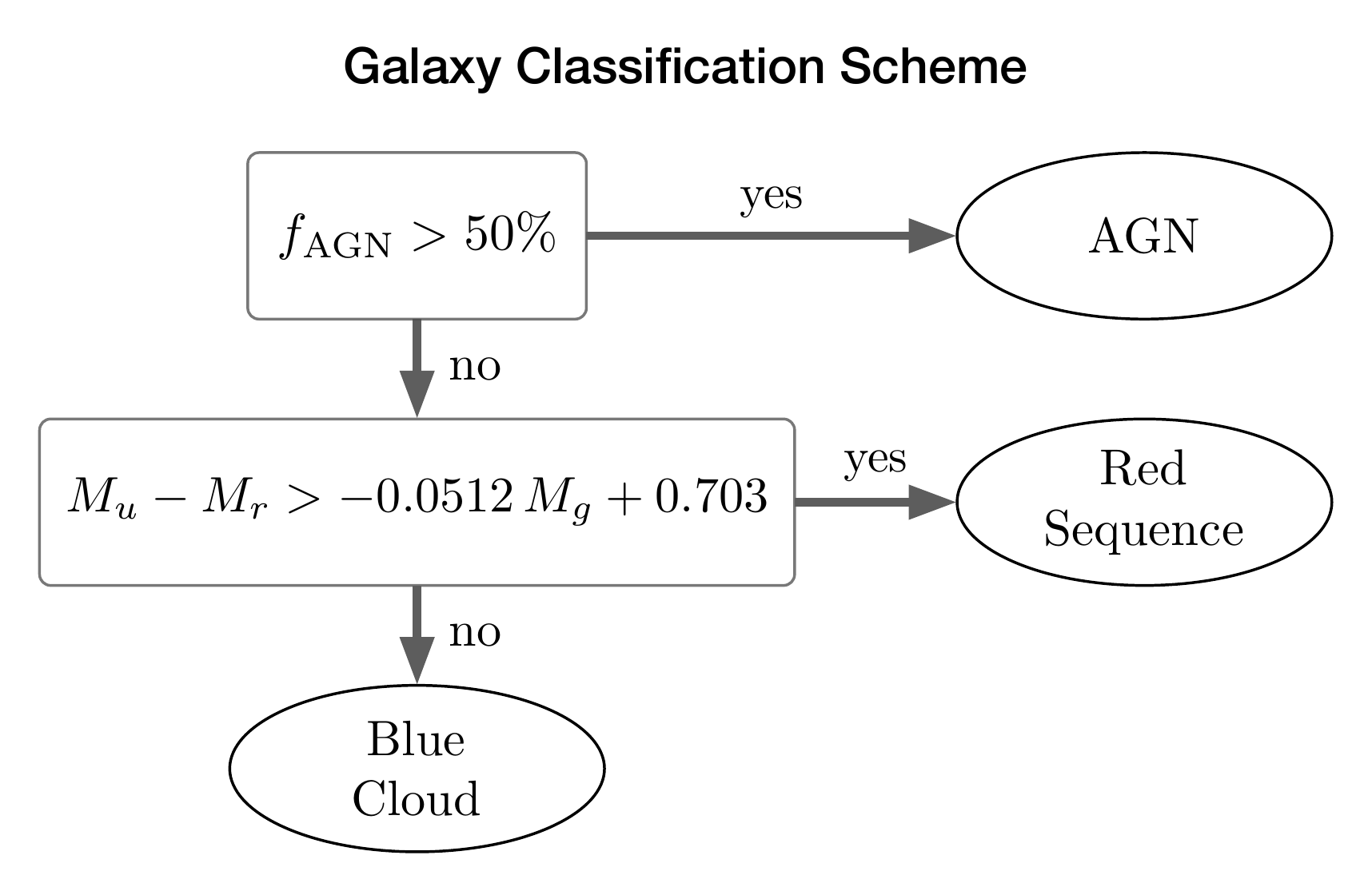}
	\end{center}
	
	\caption[Classification Flowchart]{Simple flowchart showing how galaxies were classified into AGN, red or blue galaxies in this work.}
	\label{fig:dat:class}
\end{figure}

\begin{figure*}[htb]
	\begin{center}
	\includegraphics[width=0.75\textwidth]{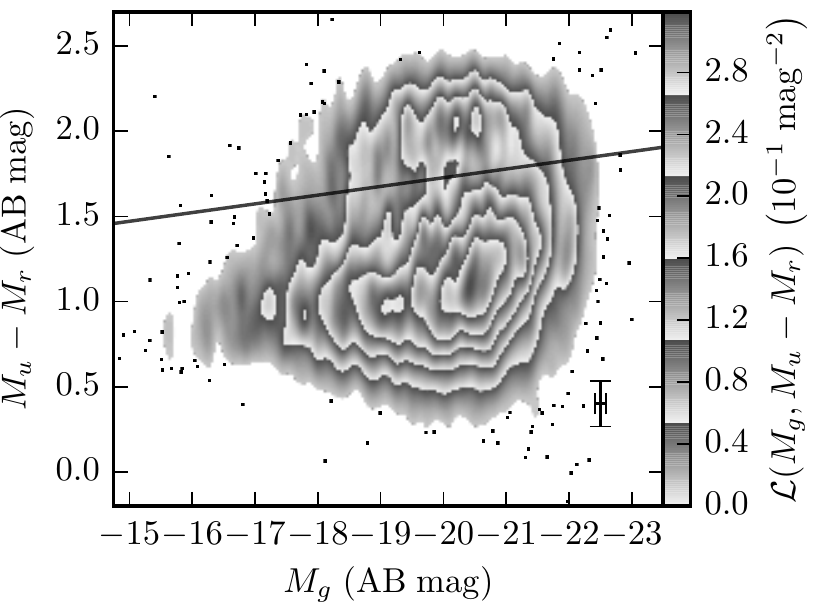}
	\end{center}
	
	\caption[Color-Magnitude Diagram]{Color-magnitude diagram showing the dividing line between the red sequence (above the line) and the blue cloud (below). The point with error bars shows the standard deviation of the smoothing kernel applied to the data in the dense region of the plot. }
	\label{fig:dat:CMD}
\end{figure*}

We also performed a classification of galaxies into three possible subsets for which SED means and covariances were measured: AGN, red galaxies, and blue galaxies. The scheme for how this classification was done is outlined in the flowchart in Figure~\ref{fig:dat:class}. The dividing line for whether a galaxy is considered an AGN is if more than 50\% of its $2.4\micron$ luminosity comes from the obscured AGN component. The dividing line for whether a galaxy is ``red" was determined empirically by examining the smoothed $M_u - M_r$ versus $M_g$, that is a standard rest frame color versus absolute magnitude diagram, shown in Figure~\ref{fig:dat:CMD}. The rest frame Sloan filter $M_u$, $M_r$, and $M_g$ were calculated by $K$-correcting observer frame Subaru $g^+$, $r^+$, and $i^+$ fluxes, respectively, from the \cite{Capak:2007} data. We experimented with a photometric classification scheme for AGN, specifically the Stern wedge from \cite{Stern:2005}, but the reduced sensitivity of the longer wavelength IRAC data meant that the blue and AGN mean SEDs were nearly the same. The final classification process shown in Figure~\ref{fig:dat:class} resulted in: $266$ AGN ($3.7\%$ of the sample), $1,906$ red sequence galaxies ($26.4\%$ of the sample), and $5,053$ blue cloud galaxies ($69.9\%$ of the sample).

\begin{deluxetable*}{cllllllrcrcc}
	\tabletypesize{\scriptsize}
	\tablewidth{\textwidth}
	\tablecaption{Excerpt of Fit Data}
	\tablehead{\colhead{\code{ID}} & \colhead{\code{ra}} & \colhead{\code{dec}} & 
		\colhead{\code{f\_Ell}} & \colhead{\code{f\_Sbc}} & \colhead{\code{f\_Irr}} & \colhead{\code{f\_AGN}} & \colhead{\code{EBmV}} & \colhead{\code{ChiSqr}} & \colhead{\code{Ndf}} &  \colhead{\code{FitMode}}
		& \colhead{\code{class}} \\
		& \colhead{$\deg$} & \colhead{$\deg$} & & & & & \colhead{mag} & & & &  }
	\startdata
		700178 & 150.305008 & 1.876265 & 0.2611 & 0.0000 & 0.4446 & 0.2943 & 0.0000 & 1.27E+02 & 11 & main & B\\
		700189 & 150.308258 & 1.916484 & 1.0000 & 0.0000 & 0.0000 & 0.0000 & 0.0000 & 8.30E+03 & 16 & alt & B\\
		700274 & 149.926743 & 1.869646 & 0.6927 & 0.0000 & 0.1911 & 0.1162 & 0.0000 & 1.98E+01 & 7 & main & B\\
		700291 & 149.890167 & 1.859292 & 0.9394 & 0.0000 & 0.0000 & 0.0606 & 0.0921 & 3.50E+02 & 13 & main & R\\
		700298 & 149.816711 & 1.916690 & 0.0803 & 0.6652 & 0.1738 & 0.0807 & 0.0000 & 1.02E+02 & 10 & main & B\\
		700447 & 150.425690 & 2.123886 & 0.1311 & 0.5809 & 0.1246 & 0.1633 & 0.0450 & 9.10E+01 & 12 & main & B
	\enddata
	\tablecomments{Excerpt from the data set included with this work in IPAC Table format. The ID column is the unique identification number given to the target in the zCOSMOS survey. \code{ra} and \code{dec} are the J2000 right ascension and declination in the zCOSMOS targets, in decimal degrees. \code{f\_Ell}, \code{f\_Sbc}, \code{f\_Irr}, and \code{f\_AGN} are the fraction of $2.4\micron$ luminosity contributed by the Elliptical, Sbc, Irregular, and obscured AGN templates, respectively. \code{EBmV}$ = (2.5 / \ln(10)) \cdot (\tau_B - \tau_V)$ is the excess extinction in the AGN obscuration model.  \code{ChiSqr} is the raw $\chi^2$ from the fitting process. \code{Ndf} is the net number of degrees of freedom in the fitting process (number of filters used minus 5), ignoring the way the effective dimensionality is altered by the constraints on the fitting process. \code{FitMode} is a character string that takes on one of two values: ``main" if $\tau_B - \tau_V$ was allowed to vary in the fitting process, ``alt" if it was set to $0$ as described in the text. \code{class} denotes the class assigned to the galaxy, and is one of `A', `R', or `B' for `AGN', `Red', and `Blue', respectively. The full table is available at: \url{https://figshare.com/articles/zCOSMOS_Template_Fractions_tbl_gz/3804210} with doi:10.6084/m9.figshare.3804210.}
	\label{tbl:dat:fitpars}
\end{deluxetable*}

\section{Results} \label{sec:results}
The template parameters for the mean SEDs, $\mu_j$ in Equations~\ref{eqn:thry:sigtmp} and the median $\tau_B - \tau_V$, of the different subsamples can be found in Table~\ref{tbl:res:meanSED}.  The template covariance matrices, $\sigma_{jk}$ in Equations~\ref{eqn:thry:sigtmp}, are in split up into Tables~\ref{tbl:res:allSEDcov}--\ref{tbl:res:BlueSEDcov} for the overall sample of, AGN, red, and blue galaxies, respectively. Using the numbers in these tables with the normalized templates and obscuration models of \cite{Assef:2010} is sufficient to calculate the covariance associated with any set of $K$-corrections. It is useful to examine graphs of the diagonal elements of the covariance, $\sqrt{\Sigma(\nu,\,\nu)}$, and the correlation function, $\rho(\nu,\,\nu') = \Sigma(\nu,\,\nu') / \sqrt{\Sigma(\nu,\,\nu)\, \Sigma(\nu',\,\nu')}$, to get a feel for how they behave, and to have as a reference for quick spectral calculations of $K$-correction covariances. 

\begin{figure*}[htb]
	\begin{center}
	\includegraphics[width=0.98\textwidth]{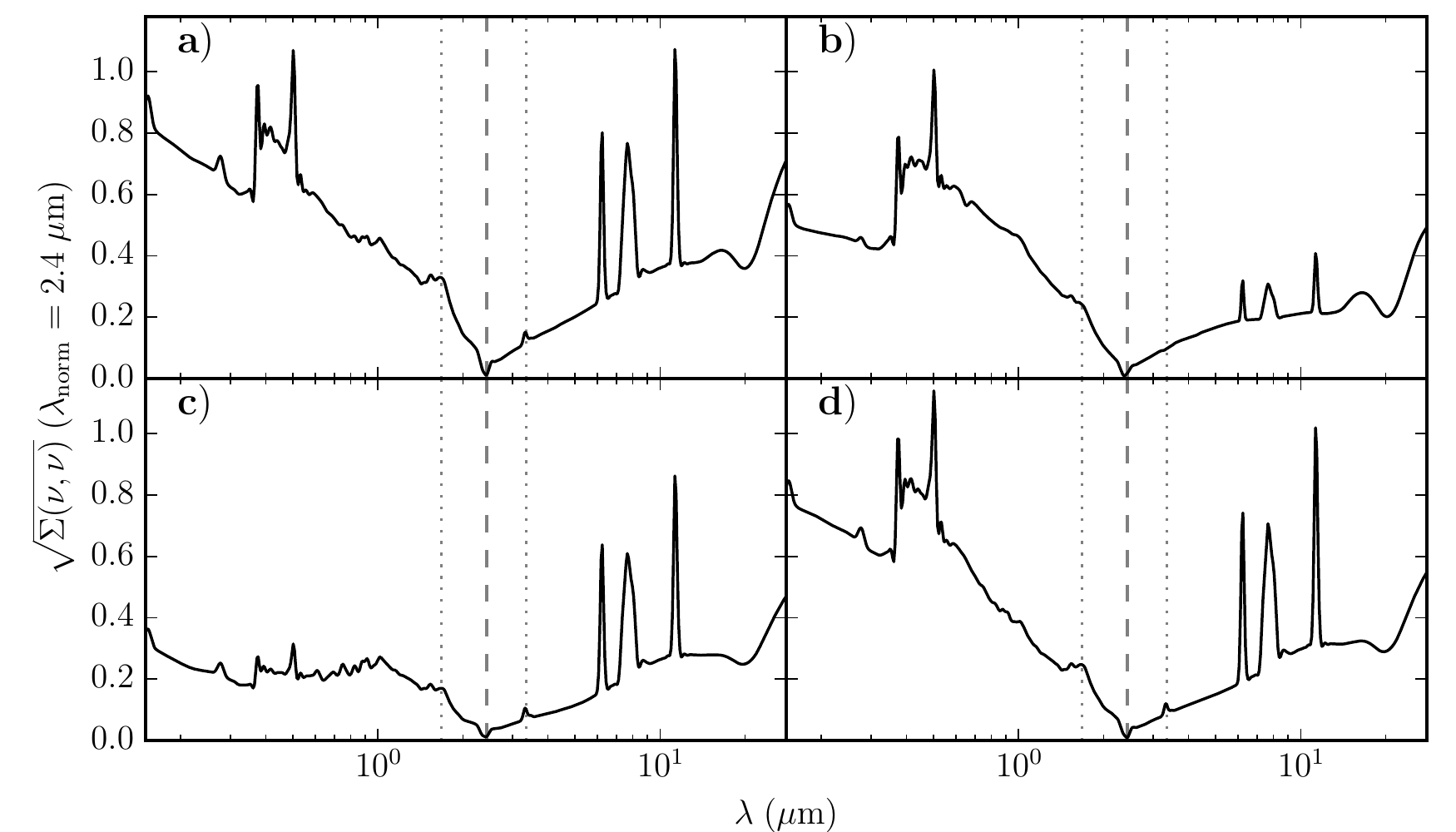}
	\end{center}
	
	\caption[SED Variance]{Panel \textbf{a} shows the SED standard deviation ($\sqrt{\Sigma(\nu,\nu)}$) for all galaxies in the sample, and Panels \textbf{b}, \textbf{c}, and \textbf{d} show the same data for AGN, red, and blue galaxies, respectively. The vertical dashed line highlights the effective wavelength of the normalization luminosity, and the dotted lines show the effective rest frame wavelength of \WISE's W1 channel for galaxies at the redshifts $z=0$ and $1$. Note the there are no units given for $\Sigma(\nu,\nu')$ because the normalization luminosity used, $L_N$, fits the standard practice in astronomy of it's weighting function, $w_N(\nu)$, having the units needed to make $L_N$ a weighted mean of $L_\nu$, that is $w_N(\nu)$ has units $[\operatorname{Hz}^{-1}]$. }
	\label{fig:res:SEDVars}
\end{figure*}

Graphs of $\sqrt{\Sigma(\nu,\,\nu)}$ for the $2.4\micron$ normalized SEDs can be found in Figure~\ref{fig:res:SEDVars}. The standard deviation increases with wavelength distance from $2.4\micron$, but there are dips and jumps around spectral features with a wide variety of strengths, specifically spectral breaks and lines. Further, the increase in the spread is steeper on the short wavelength side than the long one, supporting the assertion that simple wavelength distance is not sufficient to determine which observer frame bands are the best to $K$-correct from. The scaling on the graph is linear in $y$ and logarithmic in $x$, so the large nearly linear stretches in the graphs represent growth that is logarithmic in wavelength ratio in the standard deviation of galaxy SEDs. The final notable feature is that the spread of red galaxy SEDs, in panel~\textbf{c}, is low, as to be expected from the comparative narrowness of the red sequence in color-magnitude diagrams like Figure~\ref{fig:dat:CMD}. The comparatively large spread in AGN SEDs, panel~\textbf{b}, is surprising because the AGN selection criterion is that most of the galaxy's light at $2.4\micron$, close to the minimum of the AGN SED, comes from the AGN. This criterion explicitly limits the range of possible values for the $f_{\mathrm{AGN}}$, and implicitly limits the other fractions because they must sum to $1$. The selection criterion does affect the template covariance matrix as expected (compare the $\sigma$ column in Table~\ref{tbl:res:AGNSEDcov} to the ones in Tables~\ref{tbl:res:RedSEDcov} and \ref{tbl:res:BlueSEDcov}). The most likely culprit for the variability is how the AGN template is so different from the other three (see Figure~\ref{fig:dat:templates}). The blue cloud galaxies, in panel~\textbf{d}, have a higher spread than any of the other types of galaxies, especially in the spectral lines, other than panel~\textbf{a}, which summarizes the standard deviation for all galaxies.

\begin{figure*}[htb]
	\begin{center}
	\includegraphics[width=0.98\textwidth]{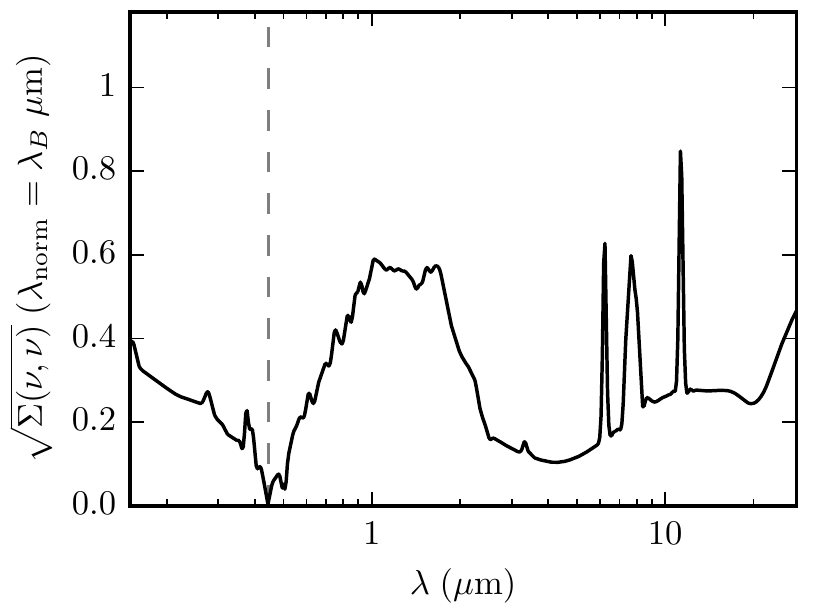}
	\end{center}
	
	\caption[SED Variance]{The SED standard deviation ($\sqrt{\Sigma(\nu,\nu)}$) for all galaxies in the sample. The vertical dashed line highlights the effective wavelength of the normalization luminosity, which is the Johnson-Cousins $B$ filter for this plot only. Note the there are no units given for $\Sigma(\nu,\nu')$ because the normalization luminosity used, $L_N$, fits the standard practice in astronomy of it's weighting function, $w_N(\nu)$, having the units needed to make $L_N$ a weighted mean of $L_\nu$, that is $w_N(\nu)$ has units $[\operatorname{Hz}^{-1}]$. }
	\label{fig:res:BVars}
\end{figure*}

The non-monotinicity of $\sqrt{\Sigma(\nu,\, \nu)}$ is actually suppressed in Figure~\ref{fig:res:SEDVars} because the normalization luminosity lies in a range of frequencies where most galaxy SEDs don't show much variety. A clearer example of the SED variance exhibiting a broad maximum can be found in Figure~\ref{fig:res:BVars}, where $\sqrt{\Sigma(\nu,\, \nu)}$ is plotted for all galaxies with a normalization luminosity in the rest frame $B$ filter instead of $2.4\micron$. The standard deviation is pinched off by the normalization near $445\nm$ and the spread among the SED templates in the $2$--$5\micron$ range is intrinsically low, producing a marked peak in the standard deviation in most of the optical and near-IR. Because the uncertainty in the $K$-correction requires input from a $B$ normalized SED and a redshift, it is not possible to say, for sure, that the variance involved in correcting from $B$ to, say, $4\micron$ is smaller than correcting from $I$ to $4\micron$. Even so, real correlations, like the far IR radio correlation, should show a pattern like this in plots of $\sqrt{\Sigma(\nu,\, \nu)}$ that cover the relevant frequency range.

\begin{figure*}[htb]
	\begin{center}
	\includegraphics[width=0.98\textwidth]{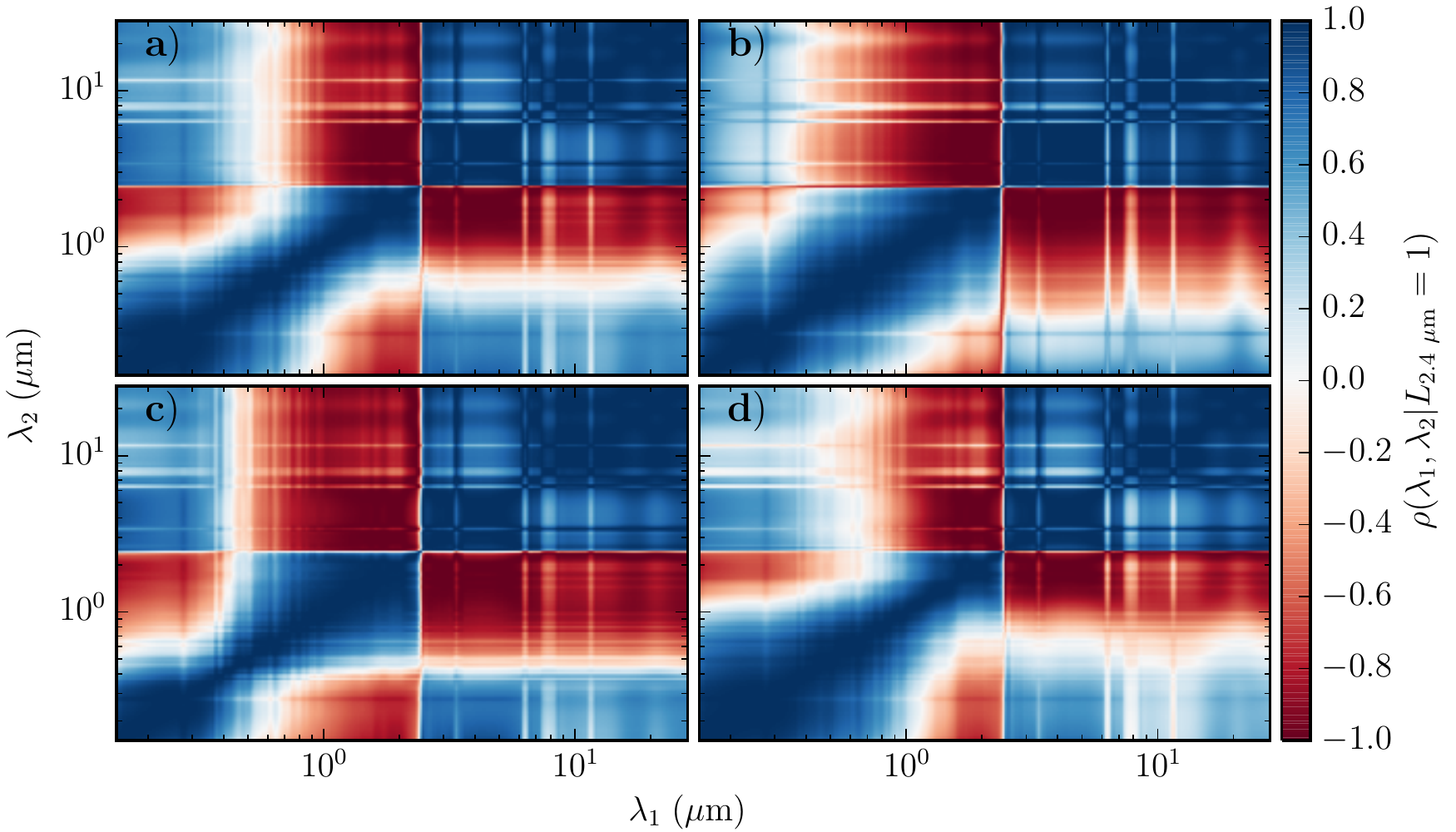}
	\end{center}
	
	\caption[SED Correlation Function]{SED correlation functions, $\rho(\nu,\,\nu') = \Sigma(\nu,\,\nu') / \sqrt{\Sigma(\nu,\,\nu) \, \Sigma(\nu',\, \nu')}$. Panel~\textbf{a} shows $\rho$ for all galaxies, while \textbf{b}, \textbf{c}, and \textbf{d} show $\rho$ for AGN, red, and blue galaxies, respectively. The dominant feature in the graphs, the sign flips across the $\lambda_{\{1,2\}} = 2.4\micron$ axes, is due to the choice of normalizing the SEDs to that wavelength. The impact of spectral features, emission and absorption lines, in the templates also stand out as horizontal and vertical striping. The final feature worth noting is the sign flip that corresponds, roughly, to when the wavelengths are on opposite sides of the point where the template SEDs that dominate the population diverge, generating the thick white lines where the sign of $\rho$ flips. }
	\label{fig:res:covar}
\end{figure*}

Very few astronomers are interested in $K$-correcting only to $2.4\micron$, and that's where the utility of the correlation function, shown in Figure~\ref{fig:res:covar}, comes in. As Equations~\ref{eqn:thry:singlvar} and \ref{eqn:thry:fullcovar} show, by combining the full $\Sigma(\nu,\, \nu')$ with the SED used in generating the $K$-correction (normalized to $2.4\micron$) and the filter curves, any covariance of $K$-corrections can be calculated. 

The most prominent features in the correlation coefficient graphs related to physics, as opposed to mathematical artifacts that comes purely from the choice of normalization wavelength, in the graphs are the thick white lines. For points on those lines, the SED colors $L_\nu(\lambda_1) / L_\nu(2.4\micron)$ and $L_\nu(\lambda_2) / L_\nu(2.4\micron)$ are uncorrelated, meaning that they contain no mutual information and, therefore, provide maximally independent information about the shape of the SED. The location of those white lines is determined by where the template SEDs that dominate the sample diverge from each other. In the case of the red galaxies (Panel~\textbf{c}) this happens roughly at the $4,000$\,\AA\ break. For the other galaxies, the diversity of SEDs is more broad and the divergence of the templates is more gradual so the main uncorrelated band is more broad and more difficult to pin down to a single phenomenon. 

The other prominent features present as horizontal and vertical striping. Those are caused by the presence of absorption and emission lines in some templates and not others. The most prominent emission lines present in the templates are: MgII ($279.8\nm$), OII (doublet, $372.6$ and $372.9\nm$), OIII (merged $495.9$ and $500.7\nm$), $\operatorname{H\alpha}$, and PAH lines at $\lambda > 3\micron$. The absorption lines are primarily a feature of the Elliptical template and that is responsible for the less prominent striping in the optical. 

\begin{deluxetable}{clllll}
	\tabletypesize{\scriptsize}
	\tablewidth{0.48\textwidth}
	\tablecaption{Mean SED Parameters}
	\tablehead{\colhead{Subsample} & \colhead{$\langle f_{\mathrm{Ell}} \rangle$} & \colhead{$\langle f_{\mathrm{Sbc}} \rangle$} & \colhead{$\langle f_{\mathrm{Irr}} \rangle$} & \colhead{$\langle f_{\mathrm{AGN}} \rangle$} & \colhead{$\overline{\tau_B - \tau_V}$\tablenotemark{a}} }
	\startdata
		all &  $0.490$ & $0.269$ & $0.114$ & $0.127$ & $0.023$ \\
		\hline
		AGN & $0.180$ & $0.076$ & $0.078$ & $0.666$ & $0.207$ \\
		Red & $0.823$ & $0.131$ & $0.011$ & $0.035$ & $0.303$ \\
		Blue & $0.380$ & $0.331$ & $0.155$ & $0.134$ & $0.015$ 
	\enddata
	\tablecomments{ Mean of the $2.4\micron$ luminosity template fractions, alongside the median excess extinction on the AGN. Numbers are given to three decimal places regardless of experimental uncertainty. }
	\tablenotetext{a}{ $\overline{\tau_B - \tau_V}$ here means the median of $\tau_B - \tau_V$. }
	\label{tbl:res:meanSED}
\end{deluxetable}

\begin{deluxetable}{lccccc}
	\tabletypesize{\scriptsize}
	\tablewidth{0.48\textwidth}
	\tablecaption{Covariance Matrix of All SED Templates}
	\tablehead{\colhead{Parameter} & \colhead{$\sigma$} & \colhead{$\delta f_{\mathrm{Ell}}$} & \colhead{$\delta f_{\mathrm{Sbc}}$} & \colhead{$\delta f_{\mathrm{Irr}}$} & \colhead{$\delta f_{\mathrm{AGN}}$} }
	\startdata
		$\delta f_{\mathrm{Ell}}$ & $0.353$ & $\hphantom{-}1.000$ & $-0.727$ & $-0.366$ & $-0.373$ \\
		$\delta f_{\mathrm{Sbc}}$ & $0.325$ & $-0.727$ & $\hphantom{-}1.000$ & $-0.209$ & $-0.224$ \\
		$\delta f_{\mathrm{Irr}}$ & $0.153$ & $-0.366$ & $-0.209$ & $\hphantom{-}1.000$ & $\hphantom{-}0.268$ \\
		$\delta f_{\mathrm{AGN}}$ & $0.163$ & $-0.373$ & $-0.224$ & $\hphantom{-}0.268$ & $\hphantom{-}1.000$
	\enddata
	\tablecomments{The $\sigma$ column contains the standard deviations of the parameters, and the rest of the columns are the correlation matrix among the template fractions. Numbers are given to three decimal places regardless of experimental uncertainty. }
	\label{tbl:res:allSEDcov}
\end{deluxetable}

\begin{deluxetable}{lccccc}
	\tabletypesize{\scriptsize}
	\tablewidth{0.48\textwidth}
	\tablecaption{Covariance Matrix of AGN SED Templates}
	\tablehead{\colhead{Parameter} & \colhead{$\sigma$} & \colhead{$\delta f_{\mathrm{Ell}}$} & \colhead{$\delta f_{\mathrm{Sbc}}$} & \colhead{$\delta f_{\mathrm{Irr}}$} & \colhead{$\delta f_{\mathrm{AGN}}$} }
	\startdata
		$\delta f_{\mathrm{Ell}}$ & $0.162$ & $\hphantom{-}1.000$ & $-0.459$ & $-0.388$ & $-0.417$ \\
		$\delta f_{\mathrm{Sbc}}$ & $0.118$ & $-0.459$ & $\hphantom{-}1.000$ & $-0.212$ & $-0.112$ \\
		$\delta f_{\mathrm{Irr}}$ & $0.136$ & $-0.388$ & $-0.212$ & $\hphantom{-}1.000$ & $-0.370$ \\
		$\delta f_{\mathrm{AGN}}$ & $0.131$ & $-0.417$ & $-0.112$ & $-0.370$ & $\hphantom{-}1.000$
	\enddata
	\tablecomments{The $\sigma$ column contains the standard deviations of the parameters, and the rest of the columns are the correlation matrix among the template fractions. Numbers are given to three decimal places regardless of experimental uncertainty. }
	\label{tbl:res:AGNSEDcov}
\end{deluxetable}

\begin{deluxetable}{lccccc}
	\tabletypesize{\scriptsize}
	\tablewidth{0.48\textwidth}
	\tablecaption{Covariance Matrix of Red SED Templates}
	\tablehead{\colhead{Parameter} & \colhead{$\sigma$} & \colhead{$\delta f_{\mathrm{Ell}}$} & \colhead{$\delta f_{\mathrm{Sbc}}$} & \colhead{$\delta f_{\mathrm{Irr}}$} & \colhead{$\delta f_{\mathrm{AGN}}$} }
	\startdata
		$\delta f_{\mathrm{Ell}}$ & $0.260$ & $\hphantom{-}1.000$ & $-0.941$ & $-0.111$ & $-0.229$ \\
		$\delta f_{\mathrm{Sbc}}$ & $0.252$ & $-0.941$ & $\hphantom{-}1.000$ & $-0.041$ & $-0.070$ \\
		$\delta f_{\mathrm{Irr}}$ & $0.043$ & $-0.111$ & $-0.041$ & $\hphantom{-}1.000$ & $-0.046$ \\
		$\delta f_{\mathrm{AGN}}$ & $0.079$ & $-0.229$ & $-0.070$ & $-0.046$ & $\hphantom{-}1.000$
	\enddata
	\tablecomments{The $\sigma$ column contains the standard deviations of the parameters, and the rest of the columns are the correlation matrix among the template fractions. Numbers are given to three decimal places regardless of experimental uncertainty. }
	\label{tbl:res:RedSEDcov}
\end{deluxetable}

\begin{deluxetable}{lccccc}
	\tabletypesize{\scriptsize}
	\tablewidth{0.48\textwidth}
	\tablecaption{Covariance Matrix of Blue SED Templates}
	\tablehead{\colhead{Parameter} & \colhead{$\sigma$} & \colhead{$\delta f_{\mathrm{Ell}}$} & \colhead{$\delta f_{\mathrm{Sbc}}$} & \colhead{$\delta f_{\mathrm{Irr}}$} & \colhead{$\delta f_{\mathrm{AGN}}$} }
	\startdata
		$\delta f_{\mathrm{Ell}}$ & $0.304$ & $\hphantom{-}1.000$ & $-0.728$ & $-0.215$ & $-0.189$ \\
		$\delta f_{\mathrm{Sbc}}$ & $0.337$ & $-0.728$ & $\hphantom{-}1.000$ & $-0.418$ & $-0.375$ \\
		$\delta f_{\mathrm{Irr}}$ & $0.162$ & $-0.215$ & $-0.418$ & $\hphantom{-}1.000$ & $\hphantom{-}0.346$ \\
		$\delta f_{\mathrm{AGN}}$ & $0.128$ & $-0.189$ & $-0.375$ & $\hphantom{-}0.346$ & $\hphantom{-}1.000$
	\enddata
	\tablecomments{The $\sigma$ column contains the standard deviations of the parameters, and the rest of the columns are the correlation matrix among the template fractions. Numbers are given to three decimal places regardless of experimental uncertainty. }
	\label{tbl:res:BlueSEDcov}
\end{deluxetable}

%

\section{Conclusion}
In this work we derived formulae for computing the uncertainty added to an observed flux when it is $K$-corrected. We also showed, by approximating the SED covariance function using template fitting data, that the choice of which observations to $K$-correct to the rest frame quantities desired should be informed by information about the variety of the SEDs of the objects in question. While the discussion in the body of this paper focused on $K$-corrections, they are just a specific type of filter transformation, and the adaptation of the formulae here to all filter transforms is trivial: just drop the factors of $(1+z)$. 

An example of a filter transformation that the covariance of observer frame SEDs can inform is the transformation from broad band filter to a spectral quantity (for example: W1 to $3.4\micron$). Note how the SED standard deviation plots in Figure~\ref{fig:res:SEDVars} have a minimum near $2.4\micron$. If the normalization luminosity were actually the spectral luminosity at $2.4\micron$ that minimum would be a zero of the function. Because the normalization is actually $^{0.38}$W1, in the notation of \cite{Blanton:2003LF} and subsequent works, the function only achieves a minimum that is close to zero at a wavelength very near $2.4\micron$. A similar plot of observer frame SED standard deviation for stars would show a similar minimum at the spectral wavelength most correlated with the broad band measurement, making that wavelength a good candidate for labeling as the filter's effective wavelength.

This study of the galaxy SED correlation function, and the mean normalized SED of galaxies, is also useful in that it feeds in to a generalization of the luminosity function that we call the spectroluminosity functional, $\Psi[L_\nu]$. We will be exploring the usefulness and mechanics of measuring $\Psi[L_\nu]$ in \cite{Lake:2016b}, and using the data from this paper and $\Psi[L_\nu]$ to measure the ordinary luminosity function, $\Phi(L)$, in \cite{Lake:2016d}.

Finally, there are definitely improvements that can be made to the techniques used here. The templates used are static, and the mean SEDs are not allowed to depend on luminosity. The latter is somewhat justified by the weak index in the power law relating $g$-band luminosity to $M_u - M_r$ color in the cut ($-0.0512$, see Figure~\ref{fig:dat:class}), but it would still be an improvement to allow for a luminosity dependence in the SED mean.

\bibliography{ThesisBib}

\end{document}